\documentclass[final]{svjour3}
\usepackage{graphicx}
\usepackage{rotating}
\usepackage{amsmath}
\usepackage{amssymb}
\usepackage{mathptmx}
\usepackage[numbers]{natbib}
\makeatletter
\journalname{Journal of Low Temperature Physics}

\bibpunct{}{}{,}{s}{}{,}

\begin{document}

\newcommand{\hdblarrow}{H\makebox[0.9ex][l]{$\downdownarrows$}-}
\title{Analytical models for the pulse shape of a superconductor-ferromagnet tunnel junction thermoelectric microcalorimeter}

\author{Z. Geng \and I. J. Maasilta}

\institute{Nanoscience Center, Department of Physics, University of Jyv{\"a}skyl{\"a},\\ Jyv{\"a}skyl{\"a}, FI-40014, Finland\\ 
\email{zhgeng@jyu.fi,maasilta@jyu.fi}}

\maketitle

\begin{abstract}
The superconductor-ferromagnet thermoelectric detector (SFTED) is a novel ultrasensitive radiation detector based on the giant thermoelectric effect in superconductor-ferromagnet tunnel junctions. We demonstrate  analytical models and solutions in the time domain for a SFTED operated as a microcalorimeter (pulse excitation), in the linear small-signal limit. Based on these solutions, the signal current and temperature pulse response were studied for two different electrical circuit models, providing design conditions for stable and non-oscillatory response.  


\end{abstract}

A superconductor-ferromagnet thermoelectric detector (SFTED) \cite{Heikkila2018} can potentially be used as a sensitive microcalorimeter to detect energetic particles and quanta such as X-rays with excellent energy resolution \cite{Chakraborty2018,Geng2020}. This type of detector is based on the giant thermoelectric effect discovered recently in superconductor-ferromagnet hybrid systems \cite{Ozaeta2014,kolenda}. Part of the novelty of such a device is that it directly transduces the absorbed energy into a measurable electrical signal without any bias power, fundamentally reducing the heat dissipation and wiring complexity demands for large sensor arrays. Here we study the time-domain signal current and temperature behavior of a SFTED operated as a microcalorimeter, with analytical models and solutions. 

The purpose of this article is twofold. On one hand, the SFTED has, up till now, only been analyzed in the frequency domain. Such frequency domain analyses are particularly useful for studies of noise and energy resolution, since stationary noise is usually uncorrelated between frequency bins in the linear, small-signal limit.  In contrast, time domain solutions provide a direct measure of the potentially complex behavior of the signal pulses, and can be used to optimize the detector and readout components. However, time-domain solutions are generally more difficult to obtain, and often only numerical solutions are available. On the other hand, although the underlining physics is drastically different, we will show that the fundamental equations of the current and temperature response of an SFTED can be cast in an analogous form to the equations for transition edge sensors (TES).  In particular, TES utilizes electrothermal feedback \cite{Irwin2005}, which has an equivalent in the SFTED through the thermoelectric coupling (Peltier effect). Modeling SFTED as an analog to the widely used and more mature TES can thus connect the studies of SFTED closer to the large body of knowledge of TESes, while helping to understand the SFTED better. 

\begin{figure}
	\centering
\includegraphics[width=0.8\linewidth]{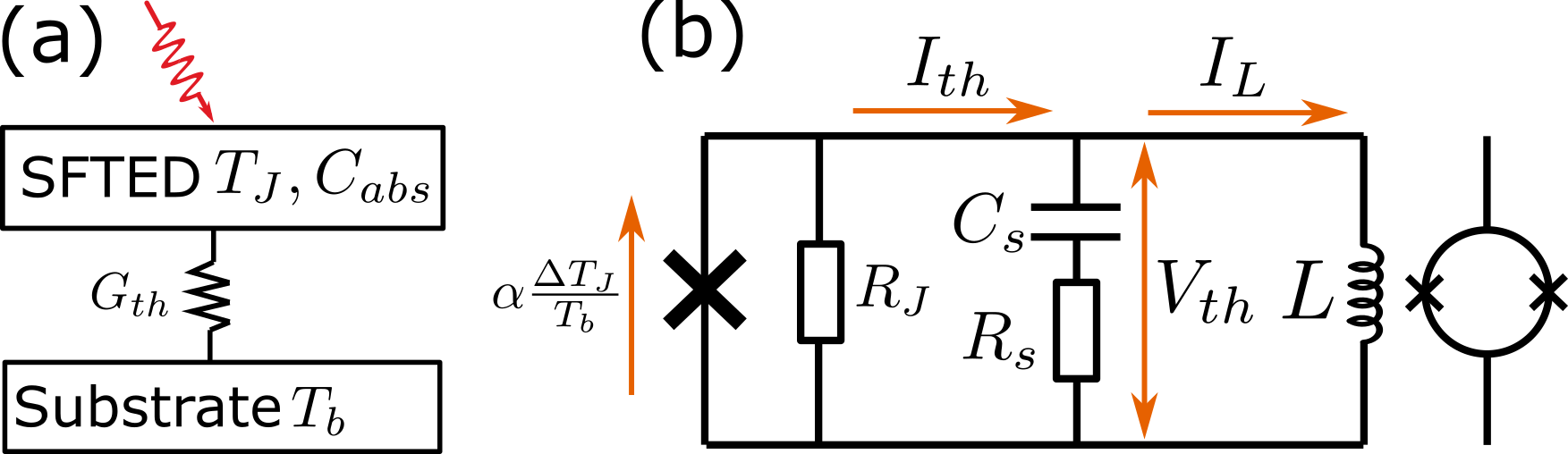}
	\caption{The schematics of the (a) the one-block thermal circuit  and (b) the electrical circuit of SFTED under the small-signal approximation, studied in this work.}
	\label{fig:circuits}
\end{figure}

In this work we study the time-domain analytical model for the simplest, one-block thermal circuit configuration shown in Fig.\ref{fig:circuits}(a). In this model, the photon absorber and the sensing electrode of SFTED are treated as a single monolithic body described by a heat capacitance $C_{abs}$ and a temperature $T_J$. This body is thermally connected to the heat bath at $T_b$ through a weak thermal link $G_{th}$, which consists of all possible heat relaxation mechanisms, including phonon  transport, electron-phonon scattering, and the thermal energy transport associated with the tunneling current itself. As low-temperature heat conduction by phonons can be engineered to a low level with membranes \cite{Irwin2005}, beams \cite{Irwin2005,Koppinen2009,Rostem2014}, phononic crystals \cite{Zen2014,Tian2019} or patterned metal features \cite{Zhang2019}, and the electron-phonon coupling is weak for a superconducting electrode \cite{Heikkila2018,Timofeev2009}, we will assume for simplicity that the tunneling current is the dominant heat relaxation channel in our numerical results. However, the analytical models presented in this article are generally applicable for any value and physical mechanism for $G_{th}$.  

The studied electrical readout circuit of the SFTED is shown in Fig.\ref{fig:circuits}(b). In this circuit, we assume a linear electrical response of the SFTED, and therefore represent the tunnel junction with an ideal current source and a junction resistance $R_J$ in parallel. The generated current from the source due to the temperature excursion of $\Delta T_J=T_J-T_b$ is $\alpha\Delta T_J/T_b$, where $\alpha$ is the thermoelectric coefficient \cite{Heikkila2018}. The current signal is designed to be inductively coupled to a SQUID readout using a large input coil with an inductance $L$ \cite{Geng2020}. In our modeling below, we first study the simplest electrical readout circuit consisting of just the inductor. Later, we also add an RC shunt, consisting of resistor $R_s$ and a capacitor $C_s$ in parallel with the SQUID input coil [Fig. 1(b)]. The understanding and optimization of such RC shunt is important because, to achieve the best energy resolution of the SFTED as an X-ray microcalorimeter, a low resistance tunneling junction (large junction area) and flux transformer coupling to the SQUID are preferred\cite{Geng2020}. The performance and stability of such a system may be hindered by resonances due to a high input inductance and parasitic capacitance. It has been shown that a proper RC shunt can damp the LC resonances in the circuit and serves as a low pass filter to further reduce the high frequency noise in a nearly noise-free manner\cite{Seppa1987,Cantor1991}, whereas a non-optimal shunt introduces excess signal loss, pulse distortion and delay, new resonances and Johnson noise, and therefore leads to performance and resolution degradation of the detector. 

In the simplest case of no RC shunt, the response of the detector can be described by two state variables: the temperature excursion $\Delta T_J$ and the current in the input coil $I_{L}=I_{th}$. These variables are governed by the electrical and thermal equations\cite{Geng2020}
\begin{equation}
	\begin{aligned}
	I_{th} &= \frac{\alpha}{T_b}\Delta T_J - \frac{1}{R_J}V_{th}\\
	\frac{d}{dt}I_L &= \frac{V_{th}}{L}	\\
	C_{abs}\frac{d}{dt}\Delta T_J & = -G_{th}\Delta T_J+\alpha V_{th},	
	\end{aligned}
\label{eq:noRC_raweqs}
\end{equation}
where $V_{th}$ is the thermoelectric voltage across the junction, as shown in Fig. 1(b). 

To highlight the similarity between the thermoelectric effect and the electrothermal feedback for a TES, we can rearrange Eq.\eqref{eq:noRC_raweqs} and present it in a matrix format, in analogy to such formulation for the TES \cite{Irwin2005}:
\begin{equation}
	\frac{d}{dt}
	\begin{pmatrix}
		I_L \\
		\Delta T_J
	\end{pmatrix}
	= 
	\begin{pmatrix}
		-\tau_{el}^{-1} & \frac{\mathcal{L}_IG_{th}}{\alpha L} \\
		-\frac{\alpha R_J}{C_{abs}} & -\tau_{I}^{-1} 
	\end{pmatrix}
	\begin{pmatrix}
		I_L \\
		\Delta T_J
	\end{pmatrix},\label{eq:noRC_eqs}
\end{equation}
where we have defined $\mathcal{L}_I=\alpha^2R_J/G_{th}T_b$ as an analog to the constant current-bias low-frequency loop gain of the TES, the electrical time constant $\tau_{el}=L/R_J$, the natural thermal time constant $\tau_{th}=C_{abs}/G_{th}$, and $\tau_{I}=\tau_{th}/(1-\mathcal{L}_I)$, the constant current thermal time constant. We note that Eq. \eqref{eq:noRC_eqs} is very similar to the one for the TES\cite{Irwin2005}: the main differences are in the definition of $\mathcal{L}_I$ and that the thermoelectric $\alpha$ appears in a dual role both within $\mathcal{L}_I$ and as the analog of the DC current of the TES. Note that the thermoelectric $\alpha$ has a unit of current, making $\mathcal{L}_I$ correctly dimensionless. It is totally different from the dimensionless logarithmic temperature sensitivity of resistance for the TES, also typically denoted by $\alpha$.

 Eqs.\eqref{eq:noRC_eqs} can be written  $d\vec{x}/dt=\pmb{D}\cdot \vec{x}$, where $\vec{x}$ is a column vector consisting of the state variables and $\pmb{D}$ is the square matrix on the RHS. General solutions are then given by $\vec{x}(t)=[I_L,\Delta T_J]^T=\sum_{n=1}^2 A_n\vec{f}_{n}\exp(-t/\tau_n)$, where $A_n$ are unitless prefactors, $\tau_n=-\lambda_n^{-1}$ the (generally) complex time constants corresponding to the eigenvalues $\lambda_n$ of the matrix $\pmb{D}$, and $\vec{f}_n$ are the associated eigenvectors, with $n=1,2$ corresponding to the two possible solutions.

The two time constants can be found to be $\tau_{\pm}^{-1}=(\tau_{el}^{-1}+\tau_{I}^{-1}\pm\sqrt{\Delta})/2$, where $\Delta=(\tau_{el}^{-1}-\tau_{I}^{-1})^2-4\mathcal{L}_I(\tau_{el}\tau_{th})^{-1}$ is the discriminator of the secular equation $||\pmb{D}\vec{f}-\lambda\vec{f}||=0$. These eigenvalues are the rise ($\tau_+$) and decay ($\tau_-$) time constants of the SFTED in response to an delta-impulse absorption event. If both time constants are positive real numbers, the SFTED is stable and has an exponential pulse decay (overdamped response), which is often the desired operation condition for microcalorimeters. A less restrictive condition for stability is that the real parts of $\tau_{\pm}^{-1}$ are positive, which allows for decaying but oscillatory solutions, as well.   

With a current readout in the overdamped case, the stability condition stated above can be shown to be always satisfied with loop gain of any value. 
However, for an unbiased SFTED the loop gain always satisfies $\mathcal{L}_I<1$, resulting from the general thermoelectric stability condition $\alpha^2R_J<G_{th}T_b$ valid for all thermoelectric systems \cite{Heikkila2018}. With that extra constraint it is easy to see that the more general stability condition is always satisfied even for the underdamped case where oscillatory solutions appear. Thus, an unbiased SFTED calorimeter is always stable.
 

\begin{figure}[h]
	\centering
	\includegraphics[width=1\linewidth]{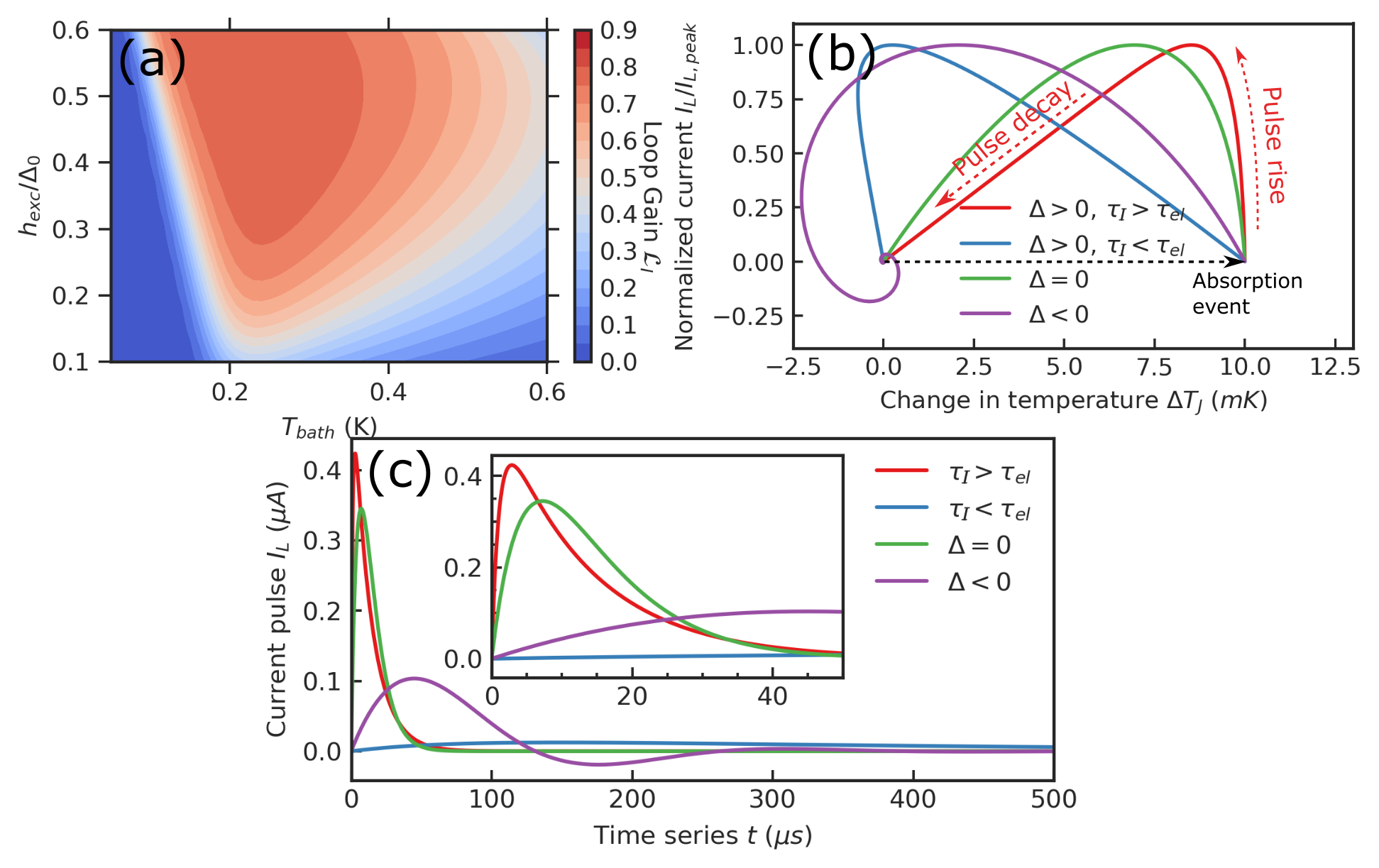}
	\caption{(a) Constant current loop gain $\mathcal{L}_I$ of the SFTED as a function of the operation temperature $T_{bath}$ and the exchange field $h_{exc}$. (b) Current-temperature pulse cycles under different operational conditions (omitting the RC shunt). The black dashed line represents the assumption of the instantaneous temperature rise in the absorber after the absorption event. (c) The signal current pulse under different operational conditions, the rising slopes of the pulses are zoomed-in in the inset. In panels (b) and (c), $\tau_I=\tau_{th}/(1-\mathcal{L}_I)$ is the constant current thermal time constant, $\tau_{el}=L/R_J$ is the electrical time constant, and fixed values $\tau_I=64\mu s$ and $R_J=10\Omega$ are used in the numerical calculations.}
	\label{fig:NoRC}
\end{figure}
In Fig.\ref{fig:NoRC}(a)-(c) we demonstrate examples of our analytical results, with different SFTED electrical time constants for the simplest model (one block thermal, inductive load). In the calculations here and later, the SFTED parameters\cite{Heikkila2018} were kept the same: Al energy gap $\Delta_{Al}(0)=0.2\,meV$, broadening parameter $\Gamma = 10^{-4} \Delta_{Al}(0)$, polarization $P=0.9$, junction normal state resistance $R_{N}=0.1\,\Omega$, absorber heat capacitance $C_{abs}=0.1\,pJ/K$ and volume of sensing electrode $V=520\,\mu m^3$. In addition, we assume that at $t=0$, the incident photon with energy $E$ is instantaneously absorbed by the absorber, raising the temperature to $T_0=E/C_{abs}+T_b$. 

In Fig.\ref{fig:NoRC}(a), the constant current loop gain $\mathcal{L}_I$ is plotted against the exchange field $h_{exc}$ and the bath temperature $T_b$. 
The exchange field is a key parameter of SFTED: the exchange interaction between the spins of magnetic ions of an insulator (\textit{e.g.} Eu$^{2+}$ in an EuS/Al/AlOx/Co device\cite{Geng2020}) and the quasiparticles in a thin superconductor (Al) at their proximity contact induces strong Zeeman splitting of the superconducting density of states\cite{Tokuyasu1988,Bergeret2018}. Combining with the spin-filtering which is applied by the ferromagnetic electrode, the electron-hole symmetry breaking is realized which leads to thermoelectric response in SFTED\cite{Heikkila2018,Ozaeta2014}. 
In the parameter space that we are interested in, the maximum $\mathcal{L}_I\approx0.8$, and high $\mathcal{L}_I$ is located in the temperature range between $0.2\,K$ to $0.4\,K$ with a relatively large exchange field. It should be also noted that the loop gain approaches zero with temperatures lower than $0.1\,K$ in general, indicating the signal becomes insensitive to temperature perturbations. For an EuS/Al bilayer, exchange field is typically about $0.4\Delta_{Al}$ without external field \cite{Strambini2017,Hao1990}, thereby from hereon, we choose to use $h_{exc}=0.4\Delta_{Al}$ and $T_b=0.23\,K$ ($\mathcal{L}_I=0.79 $) in the following calculations. 

In Fig.\ref{fig:NoRC}(b) we plot signal pulse cycles in the signal current-temperature excursion ($I_L$-$\Delta T_J$) space for four different conditions, whereas Fig.\ref{fig:NoRC}(c) shows the time evolution of the current pulses with the same conditions. The red curve ($\tau_I = 64 \mu s$,$\tau_{el}= 1 \mu s$) is at a good working point with $\Delta>0$ (overdamping) and $\tau_{I}>\tau_{el}$ to ensure the electrical circuit is fast enough to respond to the temperature change in the detector. Under these conditions, the current signal rapidly rises to its peak, with a minor decrease in the temperature before the peak is reached, and finally both current and temperature decay exponentially back to the detector's quiescent state. As a comparison, the blue curve  ($\tau_I = 64 \mu s$,$\tau_{el}= 2 ms$) shows a pulse with $\tau_{I}<\tau_{el}$. The current signal lags behind the temperature excursion, and rises slowly to the peak, whereas the temperature of detector falls back to the initial value before the decay of the current. As a result, reverse self-biasing occurs around the pulse peak, leading to a further cooling of the detector below its bath temperature. An overshoot can be observed in the temperature evolution of the detector, but is not present in the current pulse due to the slow response of the electric circuit. 

When $\Delta=0$, as shown by the green curve ($\tau_I = 64 \mu s$, $\tau_{el}= 3.8 \mu s$), Eq.\eqref{eq:noRC_eqs} has double roots, leading to equal rise and decay time constants $\tau_+=\tau_-$. Such a condition is often referred to the 'critically damped' solution, and has been considered as an optimized compromise between the energy resolution and the slew rate requirement of the readout electronics \cite{Irwin2005}. Finally, if $\Delta<0$, both current and temperature responses are oscillating, as shown by the purple curve ($\tau_I = 64 \mu s$, $\tau_{el}= 100 \mu s$), leading also to an undesired operational condition due to slow recovery. 

Adding an RC shunt to the electric circuit, as shown in Fig.\ref{fig:circuits}(b), complicates the coupled differential equations. Now the thermoelectric current $I_{th}$ is divided between the inductor $I_L$ and the shunt $I_s$, $I_{th}=I_L+I_s$, and $I_{th}$ becomes an additional state variable in addition to $I_L$ and $T_J$. By applying the equation $d(V_{th}-I_sR_s)/dt=I_s/C_s$ describing the shunt current to Eq.\eqref{eq:noRC_raweqs}, we can again rearrange and obtain a new set of governing differential equations:
\begin{equation}
	\frac{d}{dt}
	\begin{pmatrix}
		I_L \\
		\Delta T_J \\
		I_{th}
	\end{pmatrix}
	= 
	\begin{pmatrix}
		0 & \frac{\mathcal{L}_IG_{th}}{\alpha L} & -\tau_{el}^{-1} \\
		0 & -\tau_{I}^{-1} & -\frac{\alpha R_J}{C_{abs}} \\
		\frac{R_s}{R_t\tau_{RC}} 
		& \frac{\alpha}{T_bR_t}\left( \frac{R_s}{\tau_{el}}-\frac{R_J}{\tau_{I}} \right)
		& -\frac{R_J\mathcal{L}_I}{R_t\tau_{th}} - \frac{R_s}{R_t}\left(\frac{1}{\tau_{el}} + \frac{1}{\tau_{RC}}\right)
	\end{pmatrix}
	\begin{pmatrix}
		I_L \\
		\Delta T_J \\
		I_{th}
	\end{pmatrix},\label{eq:oneblock_eqs}
\end{equation}
where $R_t=R_J+R_s$ and $\tau_{RC}=R_sC_s$ is the RC time constant of the shunt. In addition, we assume the temperature rise in the absorber at $t=0$ is again a step function, which leads to a set of initial conditions $\Delta T_J(0)=\Delta T_0=E/C_{abs}$, $I_L(0) = 0$ and $I_{th}(0)=\alpha R_J\Delta T_0/R_tT_b$. 

The eigenvalues to Eq.\eqref{eq:oneblock_eqs} can be obtained by solving the non-trivial secular equation $||\pmb{D}\vec{f}-\lambda\vec{f}||=0$, which is a third order polynomial $\lambda^3+b\lambda^2+c\lambda+d=0$, with the coefficients 
\begin{equation}
	\begin{aligned}
		b &= \frac{R_s}{R_t} \left( \frac{1}{\tau_{el}} + \frac{1}{\tau_{RC}} +\frac{1}{\tau_{I}} + \frac{R_J}{R_s}\frac{1}{\tau_{th}} \right) \\
		c &= \frac{R_s}{R_t} \left( \frac{1}{\tau_{el}\tau_{RC}} + \frac{1}{\tau_{el}\tau_{th}} +\frac{1}{\tau_{RC}\tau_{I}} \right)\\
		d &= \frac{R_s}{R_t} \left( \frac{1}{\tau_{el}\tau_{RC}\tau_{th}} \right).
	\end{aligned}
\end{equation}
The three possible roots are
\begin{equation}
	\frac{1}{\tau_n}=-\lambda_n=\frac{1}{3}\bigg(b+C_n+\frac{\Delta_0}{C_n}\bigg)\quad n\in\{1,2,3\},	
	\label{eq:oneblock_delta}
\end{equation}
where $C_n=(-1/2+\sqrt{-3}/2)^{n-1}(\Delta_1/2+(-\Delta)^{1/2})^{1/3}$, $\Delta_0 = b^2-3c$, $\Delta_1 = 2b^3-9bc+27d$, and the discriminant of the polynomial equation is $\Delta = 4\Delta_0^{3} - \Delta_1^2$.

The solutions to Eqs.\eqref{eq:oneblock_eqs} are then $[I_L,\Delta T_J,I_{th}]^T=\sum_{n=1}^3 A_n\vec{f}_{n}\exp(-t/\tau_n)$, in which
\begin{equation}
		A_n\vec{f}_n =
		\begin{bmatrix}
			\frac{C_{abs}}{\alpha R_J\tau_{el}}\big(1-\frac{\tau_n}{\tau_{th}}\big) k_n \\
			k_n \\
			\frac{C_{abs}}{\alpha R_J\tau_{n}}\big(1-\frac{\tau_n}{\tau_{I}}\big) k_n
		\end{bmatrix}
\end{equation}
and
\begin{equation}
		k_n = 
		\bigg[
		\frac{\alpha R_J\tau_n}{C_{abs}}I_{th}(0)
		+ \frac{R_s}{R_t}T_0\bigg(
		\frac{\tau_n^2}{\tau_{el}\tau_{RC}} + \frac{\mathcal{L}_IR_J}{R_s}\frac{\tau_n}{\tau_{th}} + \frac{\tau_n}{\tau_{el}} + \frac{\tau_n}{\tau_{RC}} - \frac{R_t}{R_s}
		\bigg)
		\bigg]
		(d\tau_n^3-b\tau_n+2)^{-1}.	
\end{equation}

The extra root leads to an additional degree of freedom for the detector pulse behavior. However, to ensure an exponentially decaying pulse without oscillations, we can use the same condition as for the simplest model, to find proper RC-parameters for which $\Delta=0$ (critical damping). 

\begin{figure}
	\centering
	\includegraphics[width=1\linewidth]{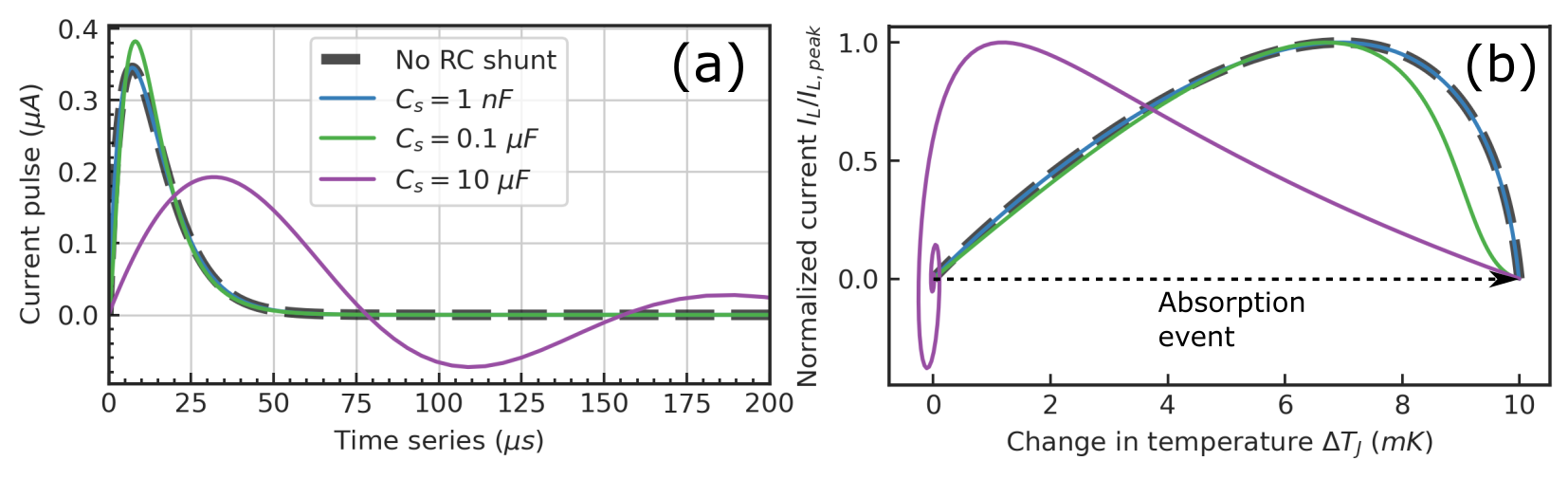}
	\caption{(a) Signal current pulses and (b) current-temperature cycles of the SFTED with different RC shunt capacitor values. All pulses have the same $R_s = 1 \Omega$.}
	\label{fig:Oneblock_model}
\end{figure}
In Fig.\ref{fig:Oneblock_model} we plot three pulses with a different shunt capacitance value $C_s$, but with the same $R_s=1\,\Omega = 10 R_{N}$, and compare them to a pulse calculated from the simplest model without a shunt. All pulses used the same detector parameters as the critically damped case in the simplest model. Under the overdamped conditions ($\Delta>0$), the high-pass RC-shunt slows down the signal current $I_L$. However, with a small enough $C_s$ (1 nF, blue curve), the rise time is dominated by $\tau_{el}$, and the pulse shape is not affected by the RC shunt. On the other hand, with the underdamped conditions $\Delta<0$ corresponding to a large $C_s$, the pole of the RC time constant is strongly interacting with both the electrical $\tau_{el}$ and the thermal time $\tau_{th}$ constants, leading to a strongly oscillating pulse in both current and temperature (purple curves). With the conditions for critical damping (green curves), the rise time of the pulse is slowed somewhat, but the peak of the pulse is amplified, resulting in an ideal operational condition for the detector.
 
In conclusion, we have reformulated the coupled differential equations of an unbiased superconductor-ferromagnet thermolectric detector (SFTED) with a one-block thermal model and an inductive current readout, with and without an additional RC shunt circuit. These equations were written in analogy to the time-domain equations of transition edge sensors (TES), to gain better understanding of the device. Based on the analytical solutions of these equations for a pulse excitation (calorimetry), the signal current and temperature response of the SFTED has been studied. In particular, the design conditions for a stable and non-oscillatory response have been given and discussed. Following the approach demonstrated here,  design and optimization conditions for  more complicated electrical and thermal models could be obtained straightforwardly in the future.

\begin{acknowledgements}
This study was supported by the Academy of Finland Project Number 341823 and by the European Union’s Horizon 2020 research 
and innovation programme under grant agreement No 800923 (SUPERTED). We thank T. T. Heikkil\"a for discussions.
\end{acknowledgements}



\begin{thebibliography}{99}


\bibitem{Heikkila2018}
T. T. Heikkil{\"a}, R. Ojajärvi, I. J. Maasilta, E. Strambini, F. Giazotto, and F. S. Bergeret, {\it Phys. Rev. Appl.} \textbf{10}, (2018). DOI:10.1103/PhysRevApplied.10.034053

\bibitem{Chakraborty2018}
S. Chakraborty and T. T. Heikkil{\"a}, {\it J. Appl. Phys.} \textbf{124}, 123902 (2018). DOI:10.1063/1.5037405

\bibitem{Geng2020}
Z. Geng, A.P. Helenius, T.T. Heikkil{\"a}, and I.J. Maasilta, {\it J. Low Temp. Phys.} \textbf{99}, 585 (2020). DOI:10.1007/s10909-020-02419-0

\bibitem{Ozaeta2014}
A. Ozaeta, P. Virtanen, F. S. Bergeret, and T. T. Heikkil{\"a}, {\it Phys. Rev. Lett.} \textbf{112}, 057001 (2014). DOI:10.1103/PhysRevLett.112.057001

\bibitem{kolenda} 
S. Kolenda, M. J. Wolf, and D. Beckmann,  {\it Phys. Rev. Lett.} \textbf{116}, 097001 (2016). DOI:0.1103/PhysRevLett.116.097001

\bibitem{Irwin2005}
K. Irwin and G. Hilton, in {\it Cryogenic Particle Detection}, Ed. Ch. Enss, Springer, Heidelberg (2005). DOI:10.1007/10933596\_3

\bibitem{Koppinen2009}
P. J. Koppinen and I. J. Maasilta, {\it Phys. Rev. Lett.} \textbf{102}, 165502 (2009). DOI:10.1103/PhysRevLett.102.165502

\bibitem{Rostem2014}
K. Rostem, D. T. Chuss, F. A. Colazo, E. J. Crowe, K. L. Denis, N. P. Lourie, S. H. Moseley, T. R. Stevenson, and E. J. Wollack, {\it J. Appl. Phys.} \textbf{115}, 124508 (2014). DOI:10.1063/1.4869737 

\bibitem{Zen2014}
N. Zen, T.A. Puurtinen, T.J. Isotalo, S. Chaudhuri, and I.J. Maasilta, {\it Nat. Commun.} \textbf{5}, 3435 (2014). DOI:10.1038/ncomms4435	

\bibitem{Tian2019}
Y. Tian, T. A. Puurtinen, Z. Geng, and I. J. Maasilta, {\it Phys. Rev. Applied} \textbf{12}, 014008 (2019). DOI:10.1103/PhysRevApplied.12.014008

\bibitem{Zhang2019}
X. Zhang, S. M. Duff, G. C. Hilton, P. J. Lowell, K. M. Morgan, D. R. Schmidt, and J. N. Ullom
{\it Appl. Phys. Lett.} \textbf{115}, 052601 (2019). DOI:10.1063/1.5097173

\bibitem{Timofeev2009}
A. V. Timofeev, C. Pascual García, N. B. Kopnin, A. M. Savin, M. Meschke, F. Giazotto, and J. P. Pekola,
{\it Phys. Rev. Lett.} \textbf{102}, 017003 (2009). DOI:10.1103/PhysRevLett.102.017003

\bibitem{Seppa1987}
H. Sepp{\"a} and T. Ryh{\"a}nen, {\it IEEE Trans. Magn.} \textbf{23}, 1083 (1987), DOI: 10.1109/TMAG.1987.1065125.

\bibitem{Cantor1991}
R. Cantor, T. Ryh\"anen, D. Drung, H. Koch and H. Sepp{\"a}, {\it IEEE Trans. Magn.} \textbf{27}, 2927 (1991). DOI: 10.1109/20.133822.

\bibitem{Tokuyasu1988}
T. Tokuyasu, J.A. Sauls, and D. Rainer, {\it Phys. Rev. B} \textbf{38}, 8823 (1988). DOI: 10.1103/PhysRevB.38.8823

\bibitem{Bergeret2018}
F. S. Bergeret, M. Silaev, P. Virtanen, and T. T. Heikkil{\"a}, {\it Rev. Mod. Phys.} \textbf{90}, 041001 (2018). DOI: 10.1103/RevModPhys.90.041001

\bibitem{Strambini2017}
E. Strambini, V.N. Golovach, G. De Simoni, J.S. Moodera, F.S. Bergeret, and F. Giazotto, {\it Phys. Rev. Mater.} \textbf{1}, 1 (2017). DOI: 10.1103/PhysRevMaterials.1.054402

\bibitem{Hao1990}
X. Hao, J.S. Moodera, and R. Meservey, {\it Phys. Rev. B} \textbf{42}, 8235 (1990). DOI: 10.1103/PhysRevB.42.8235





\end{thebibliography}
\end{document}